\documentclass[conference]{IEEEtran}
\IEEEoverridecommandlockouts

\usepackage{cite}
\usepackage{amsmath,amssymb,amsfonts}
\usepackage{graphicx}
\usepackage{textcomp}
\usepackage{xcolor}
\def\BibTeX{{\rm B\kern-.05em{\sc i\kern-.025em b}\kern-.08em
    T\kern-.1667em\lower.7ex\hbox{E}\kern-.125emX}}
\usepackage{wrapfig} 
\usepackage{orcidlink}
\usepackage{times}
\usepackage{amsmath}
\usepackage{amssymb}
\usepackage{amsfonts}
\usepackage{xspace}
\usepackage{textcomp}
\usepackage{stfloats}
\usepackage{verbatim}
\usepackage{graphicx}
\usepackage[inline]{enumitem}
\usepackage{titlesec}
\usepackage{todonotes}
\presetkeys{todonotes}{inline,backgroundcolor=yellow,caption={}}{}
\graphicspath{ {./images/} }
\usepackage[export]{adjustbox}

\usepackage{stackengine,xcolor}
\usepackage{cite}
\usepackage{xspace}

\usepackage{algorithm}
\usepackage[noend]{algpseudocode}
\usepackage{algorithmicx}
\usepackage{setspace}
\usepackage{listings}
\usepackage{xcolor}

\lstdefinelanguage{Solidity}{
  keywords={
    pragma, solidity, contract, function, return, uint256, uint, public, private, internal, external, view, pure, emit, require, modifier, new, assembly, fallback, receive, delete, pay, address, mapping, struct, enum, event, require, assert, revert, fallback, receive, payable, bytes, uint8, uint16, uint32, uint64, uint128, uint256, int8, int16, int32, int64, int128, int256, string, bytes32, bytes, returns,
  },
  keywordstyle=\color{blue}\bfseries\small,
  ndkeywords={
    struct, mapping, event, enum, address, require, revert, assert, payable, memory, storage, string, bytes, fallback, receive, delete
  },
  ndkeywordstyle=\color{purple}\bfseries\small,
  identifierstyle=\color{black}\small,
  sensitive=true,
  comment=[l]{//},
  morecomment=[s]{/*}{*/},
  commentstyle=\color{gray}\ttfamily\small,
  stringstyle=\color{orange}\ttfamily\small,
  morestring=[b]',
  morestring=[b]",
  tabsize=2,
  showspaces=false,
  showstringspaces=false
}

\PassOptionsToPackage{hyphens}{url}
\usepackage{hyperref}
\usepackage{xspace}

\def\sec{Section\xspace}

\definecolor{procColor}{HTML}{2ECC71}
\usepackage{multirow}
\usepackage{comment}

\usepackage[T1]{fontenc}
\usepackage[numbers,sort&compress]{natbib}
\usepackage[normalem]{ulem}
\usepackage{tablefootnote}
\usepackage{lipsum}
\usepackage{threeparttable}

\newlist{inlinelist}{enumerate*}{1}
\setlist[inlinelist]{label=(\arabic*), itemjoin={{, }}, itemjoin*={{, and }}}
\usepackage{float}

\makeatletter
\renewcommand{\ALG@beginalgorithmic}{\small}
\renewcommand{\algorithmicindent}{1em}
\makeatother

\usepackage[noend]{algpseudocode}
\algrenewcommand\algorithmicindent{1em}
\newcommand{\Continue}{\State \textbf{continue}}

\begin{document}
\title{
Static Analysis for Detecting Transaction Conflicts in Ethereum Smart Contracts\\
}
\author{
  \IEEEauthorblockN{Atefeh Zareh Chahoki\IEEEauthorrefmark{1}
                    \qquad
                    Marco Roveri\IEEEauthorrefmark{1}}%
  \IEEEauthorblockA{\IEEEauthorrefmark{1}%
    Department of Information Engineering and Computer Science, University of Trento, Italy\\
    \{atefeh.zareh, marco.roveri\}@unitn.it}
}

\maketitle

\begin{abstract}

Ethereum smart contracts operate in a concurrent environment where multiple transactions can be submitted simultaneously. However, the Ethereum Virtual Machine (EVM) enforces sequential execution of transactions within each block to prevent conflicts arising from concurrent access to the same state variables. Although this approach guarantees correct behavior, it limits the ability of validators to leverage multi-core architectures for faster transaction processing, thus restricting throughput. Existing solutions introduce concurrency by allowing simultaneous transaction execution combined with runtime conflict detection and rollback mechanisms to maintain correctness. However, these methods incur significant overhead due to continuous conflict tracking and transaction reversion. Recently, alternative approaches have emerged that aim to predict conflicts statically, before execution, by analyzing smart contract code for potential transaction interactions. Despite their promise, there is a lack of comprehensive studies that examine static conflict detection and its broader implications in specific smart contracts. This paper fills this important gap by proposing a novel static analysis method to detect potential transaction conflicts in Ethereum smart contracts. Our method identifies read-write, write-write, and function call conflicts between transaction pairs by analyzing state variable access patterns in Solidity contracts. We implement a tool that parses contract code and performs conflict detection. Evaluation on a dataset of real-world Ethereum smart contracts demonstrates that our approach achieves high precision in identifying potential conflicts. By enabling proactive conflict detection, our tool supports further design of transaction scheduling strategies that reduce runtime failures, enhance validator throughput, and contribute to blockchain scalability.
\end{abstract}

\begin{IEEEkeywords}
Blockchain, Ethereum, smart contracts, static analysis, transaction conflicts, concurrency.
\end{IEEEkeywords}

\section{Introduction}
Ethereum smart contracts are autonomous programs deployed on the Ethereum blockchain that facilitate decentralized and trustless interactions between parties. These contracts are commonly written in high-level languages such as Solidity and compiled to bytecode executed by the Ethereum Virtual Machine (EVM). To maintain determinism and avoid inconsistent state updates, the current EVM enforces strict sequential execution of transactions within each block by validators. In two distinct aspects of blockchain security and scalability, there is a shared gap that motivated the current research, which we elaborate on in the following.

\textbf{Security.} Due to the immutability essence of smart contracts on the blockchain, the endurance of the correctness of the transactions is crucial. Security vulnerabilities in smart contracts can stem from low-level bugs (e.g., arithmetic overflows), misuse of control flow constructs (e.g., reentrancy), or more nuanced issues involving transaction-level interactions. To address such problems, researchers have employed various analysis techniques: \emph{fuzzy analysis}, which tests contracts by mutating input values to trigger unexpected behaviors; \emph{dynamic analysis}, which monitors contracts during execution to detect anomalies; and \emph{static analysis}, which examines the contract code without execution. Among these auditing approaches, static analysis is the focus of this research. Tools such as Oyente~\cite{Luu_Making_2016}, Mythril~\cite{smashing_mueller_2018}, Slither~\cite{Feist_Slither_2019}, and Securify~\cite{Tsankov_Securify_2018} have proven to be effective in detecting known vulnerability patterns. More recently, advanced techniques have improved the precision of static analysis, incorporating control flow graphs, data flow summaries, and interprocedural analyses to detect reentry, timestamp dependency, and gas-related issues. However, these tools generally lack the ability to identify higher-order transaction-level conflicts, which is the focus of this study.

\textbf{Scalability.} Ethereum faces substantial throughput limitations due to its inherently sequential transaction model. Runtime solutions, such as optimistic concurrency control and speculative execution frameworks~\cite{Adding_Dickerson_2020}, execute transactions in parallel and rollback upon detecting conflicts. However, these approaches suffer performance degradation as conflict rates increase. Conthereum~\cite{Conthereum_Zareh_2025} addresses this shortcoming by scheduling transactions using a preventive strategy. In this solution, the conflicts between transactions are used prior to execution, enabling validators to generate conflict-free schedules. The success of this architecture hinges on the accuracy and completeness of the conflict information provided.
While Conthereum demonstrates the value of static conflict information in performance optimization, the broader scalability context demands an analysis of this type. As Layer 2 rollups and sharding approaches evolve, many rely on batch reordering or parallel execution schemes. These systems also require early identification of transaction interferences to ensure determinism, maximize throughput, and minimize gas costs. Therefore, a dedicated static conflict analyzer is crucial for both the security and the broader scalability of the Ethereum network.

The reliable and secure execution of Ethereum smart contracts is critically dependent on understanding how transactions interact with shared contract state. Although the EVM guarantees atomicity and sequential execution of transactions within a block, conflicts between transactions, arising from concurrent access to shared state variables, remain a latent source of vulnerabilities and performance bottlenecks. In particular, transaction ordering dependence, front-running risks, and unpredictable execution outcomes are often a consequence of implicit inter-transaction conflicts. 

This paper makes the following contributions.

First, we present a novel static analysis framework that extracts conflict patterns from smart contract code and identifies potential transaction conflicts without requiring execution traces. Unlike dynamic approaches, our method provides exhaustive coverage of all syntactically reachable state interactions, allowing early detection of conflict-prone function pairs. This is essential not only for optimizing transaction scheduling—such as in parallel execution engines or rollup batch reordering—but also for improving contract auditability, refactorability, and formal verification to focus on more conflicting and error prone functions. Moreover, by statically identifying state-dependent interactions, the analysis supports risk-aware prioritization of test cases and highlights ordering-sensitive logic that may otherwise be exploited through Miner Extractable Value (MEV) strategies. Thus, our work offers a foundational tool for enhancing the robustness, efficiency, and security of smart contract-based systems at scale.

Second, we provide a novel conflict taxonomy that covers both direct and indirect conflict types. We propose the following classification:

\begin{itemize}
    \item \textbf{Direct Access Conflicts}: Conflicts that arise when transactions directly access the same state variable, including:
    \begin{itemize}
        \item \textbf{Read-Write (RW)}: One transaction reads a variable while another writes to it.
        \item \textbf{Write-Write (WW)}: Both transactions write to the same variable.
    \end{itemize}
    
    \item \textbf{Transitive Access Conflicts}: Conflicts caused by indirect state access through nested or chained function calls, where one transaction invokes a function that modifies or reads state variables accessed by another transaction.

\end{itemize}
The Solidity code can thus be inspected to identify all pairs of functions that may access overlapping contract states under various access modes.

Third, we implement our approach in a tool that parses Solidity smart contracts, analyzes state variable access patterns, and detects potential conflicts with severity ratings. The tool generates comprehensive reports and visualizations to help users understand the identified conflicts and facilitate their utilization in their domain.

Fourth, we evaluated the proposed framework on a dataset of real-world Ethereum smart contracts. The results show that our approach effectively identifies potential conflicts with high precision. The results indicate that read-write conflicts are the most common, followed by write-write conflicts and function call conflicts.

Finally, we provide a proof of zero false negatives and low false positives, ensuring the reliability of the analysis in safety-critical contexts such as transaction scheduling.

The remainder of this paper is organized as follows. \sec~\ref{sec:Background} provides essential background on smart contract execution and analysis techniques. \sec~\ref{sec:solution} details our proposed static analysis approach. \sec~\ref{sec:Implementation} describes the implementation of our tool. Experimental results and analysis are presented in \sec~\ref{sec:Evaluation}. We compare our work with existing research in \sec~\ref{sec:RelatedWorks} and conclude in \sec~\ref{sec:Conclusion} with potential directions for future work.

\section{Background}~\label{sec:Background}

\subsection{Smart contracts}\label{ssec:Background_Smart_Contract}
The \textit{blockchain} is a foundational design pattern to facilitate pure peer-to-peer distributed computation. Initially implemented by Bitcoin in 2009 as a cryptocurrency~\cite{nakamoto_bitcoin_2008}, it has since evolved into a robust infrastructure capable of executing a wide range of generalized functionalities beyond the cryptocurrency. Ethereum~\cite{buterin_ethereum_white_2013}, in particular, pioneered this expansion in \citeyear{buterin_ethereum_white_2013} by introducing the \textit{Ethereum Virtual Machine (EVM)}, a Turing-complete state machine that handles both the deployment and execution of codified arbitrary business logic scripts named \textit{smart contracts}. All blockchain networks maintain securely shared data named \textit{ledger} through a \textit{consensus protocol} facilitated by volunteer operators named \textit{miners} and \textit{validators}. The transactions announced by users are initially stored in the \textit{mempool} (short for "memory pool"), a temporary storage area where pending transactions wait to be included in the next block. Miners select transactions from the mempool that offer higher fees first, as this maximizes their rewards and persists them into the next block if the transaction is valid. This validity check includes the correctness of transaction format, having a sufficient gas fee, nonce verification, signature verification, balance check, double-spending prevention, compliance with consensus rules, and, regarding smart contracts, ensuring that the transaction can execute without error. Then the miners extend the blockchain by one block in each \textit{block interval}, which varies between different networks such as 10 minutes in Bitcoin or 12 seconds in Ethereum, where miners compete according to the specifications of the consensus protocol. Smart contracts introduce the concept of \textit{gas}, irrespective of the consensus protocol they utilize. Gas represents the computational effort and cost required to execute operations within smart contracts to compensate for the resources used in the execution of smart contracts or transactions. It prevents infinite loops, restricts resource consumption, and maintains the efficiency and security of the network. Smart contracts are written in various programming languages that are tailored to specific blockchain platforms. Solidity is predominant in Ethereum, Vyper aims for simplicity and security, Michelson is used on Tezos, and C was developed for Diem (formerly known as Libra). Solidity is the dominant language for smart contracts, and the remainder of this subsection presents the minimal technical information required in this study.

\subsubsection{Function Visibility}\label{ssec:Background-Function-Visibility}
Solidity has four function visibility specifiers: public functions callable by any smart contract, private functions callable only within the current contract, internal functions callable by the current and derived contracts, and external functions callable only by external contracts, which are more gas-efficient than public functions. For brevity, we use, in the remainder of this document, the notation $C.Func.Public$ to refer to all public functions of a smart contract $C$, and similarly to the other three visibilities of functions.

\subsubsection{Function State Mutability}
In Solidity 0.5.x, functions can have one of three types of state mutability:
\begin{enumerate*}[label=\alph*)]
    \item \textbf{Pure}: The function does not read or write any data.
    \item \textbf{View}: The function reads some data from the contract's storage, but it does not modify anything.
    \item \textbf{Non-reentrant} and \textbf{non-payable} (default): The function can read and write from/to the contract's storage.
\end{enumerate*}

\subsubsection{Variable Visibility}\label{ssec:Background-Variable-Visibility}
In Ethereum smart contracts, the visibility of the variable determines the level of access of the state variables. \texttt{Public} variables can be accessed both internally within the contract and externally by other contracts or users, with Solidity automatically generating getter functions for them. \texttt{Private} variables are only accessible within the contract in which they are defined and are not visible to derived contracts or external entities. \texttt{Internal} variables can be accessed within the contract and by derived contracts, but not by external callers. Despite their visibility, all state variables, irrespective of their access level, are stored on the blockchain and persist across transactions.

\subsubsection{Memory Management in Ethereum Smart Contracts}\label{ssec:background-memory}
The EVM utilizes various types of memory: memory, calldata, stack, storage, logs, and EVM code. Each serves a specific purpose and has a distinct cost.

To illustrate these concepts, we introduce the simple smart contract in the Listing~\ref{lst:solidity-example-code}.

\begin{lstlisting}[language=Solidity, caption={Example Solidity Contract}, label={lst:solidity-example-code}, breaklines=true, breakatwhitespace=true,float=t!]
pragma solidity ^0.8.0;

contract Example {
  uint256[] public storageArray;
  uint256 private storageSize;

  // Define an event
  event StorageValueAdded(uint256 indexed value);

  function addToStorage(uint256 _value) public {
    storageArray.push(_value);
    storageSize++; // Increment the private variable
    // Emit the event
    emit StorageValueAdded(_value);
  }

  function addToMemory(uint256 _value) public view returns (uint256[] memory) {
    uint256[] memory memoryArray = new uint256[](10);
    memoryArray[0] = _value;
    return memoryArray;
  }
}
\end{lstlisting}

\begin{description}[ labelindent=0pt, labelsep=0.5em, leftmargin=1em]
\item[Memory,] also known as \textit{transient memory}, is a temporary storage that exists only during the lifetime of a function. Unlike storage, memory is reset after each transaction, making it more cost-effective. In the example code shown in Listing \ref{lst:solidity-example-code}, the function \texttt{addToMemory} creates a \texttt{memoryArray} in \textit{memory}, which only lasts for the duration of the function execution. Due to its significantly lower cost compared to storage, memory is ideal for temporary computations within a single transaction.

\item[Calldata] is a temporary data storage location used to hold function arguments passed from an external caller, such as a user or another contract. Unlike memory, calldata is read-only and cannot be modified by the function. Therefore, if you need to modify a function argument, you must first copy it into memory. Calldata is beneficial for passing large amounts of data to a function without incurring the high gas costs associated with copying data into memory. By using calldata, you can reduce gas consumption and avoid the overhead of data copying. In the Solidity example shown in Listing \ref{lst:solidity-example-code}, the parameter \texttt{\_value} in both function signatures is stored in calldata.

\item[Stack] is a collection of "words" or "slots," each 256 bits long, with a maximum capacity of 1024 items. The EVM operates as a stack machine, using this Last In, First Out (LIFO) structure to handle opcode input arguments and store results. In the Solidity example in Listing~\ref{lst:solidity-example-code}, when the \texttt{addToStorage(uint256 \_value)} function is called, the value \_value is pushed onto the stack. The EVM then uses this value to update \texttt{storageArray} by calling \texttt{push}, which involves interacting with the storage layer. This demonstrates how values are temporarily managed on the stack while interacting with persistent storage.

\item[Storage] refers to the persistent memory on the blockchain that maintains state across transactions. It holds contract's \textit{state variables} (i.e., variables declared within a contract but outside any function) and exists for the entire lifecycle of the contract. The concept of storage memory in blockchains is unique because, with smart contracts, the data is protected by the cryptographic sealing that blockchains provide, ensuring its tamper-proof nature. In contrast, in other programming environments, long-term data storage is typically managed through file systems or databases.

Each contract has its own read-write storage area, divided into ${2^{256}}$ slots of 32 bytes each, indexed from 0 to ${2^{256}-1}$. All slots are initialized to zero. Given the permanence and the large number of slots, storage is considered expensive due to the high gas costs associated with reading and writing operations.

As discussed in \ref{ssec:Background-Variable-Visibility}, all variables, regardless of their visibility, are stored on the blockchain. In the Solidity example shown in Listing \ref{lst:solidity-example-code}, both the public variable \texttt{storageArray} and the private variable \texttt{storageSize} are state variables stored in \textit{storage}, meaning they persist across function calls and transactions. The \texttt{addToStorage} function, which appends values to \texttt{storageArray}, incurs significant gas costs due to the storage operations involved.

\item[Logs] are a mechanism in the EVM for recording events that occur during the execution of smart contracts. Contracts emit events to provide external visibility into specific actions or state changes without modifying the contract’s storage. This is done using the \texttt{emit} keyword followed by an event definition, as demonstrated in the example contract \ref{lst:solidity-example-code} \texttt{addToStorage} function. Logs are stored in a separate structure known as the "\textit{event log}," which is more gas-efficient than storing data directly on-chain. Logs can include indexed parameters, making it easier for external applications to filter and search for specific events. This feature is particularly useful for decentralized applications that need to track contract interactions or trigger off-chain processes based on contract activity.

\item[EVM Code] is one of the key memory types in the EVM. It represents the low-level bytecode that the EVM executes to perform smart contract operations. Unlike other memory types, such as stack and memory, EVM code is not used for storing data but for defining the sequence of operations that execute smart contract logic. This bytecode is generated from high-level languages like Solidity and consists of a series of opcodes, each with a specific gas cost that affects the execution cost.

\end{description}

\section{Related Works}\label{sec:RelatedWorks}
Research on smart contract analysis has focused primarily on detecting security vulnerabilities rather than transaction conflicts. In this section, we review existing work on smart contract analysis and highlight the gaps that our approach addresses.

\subsection{Smart Contract Vulnerability Detection}
Several tools have been developed to detect security vulnerabilities in smart contracts. Oyente \cite{Luu_Making_2016} was one of the first tools to use symbolic execution to detect vulnerabilities such as reentrancy, timestamp dependence, and mishandled exceptions. Mythril \cite{smashing_mueller_2018} extends this approach with control flow analysis and taint tracking. Securify \cite{Tsankov_Securify_2018} uses formal verification to verify compliance with security properties. These tools focus on detecting known vulnerability patterns rather than transaction conflicts.

\subsection{Concurrency in Smart Contracts}
Kolluri et al. \cite{kolluri2019exploiting} demonstrate how transaction-ordering dependencies can be exploited in decentralized exchanges. Chakravarty et al. \cite{chakravarty2020utxo} propose the UTXO model as an alternative to the Ethereum account-based model to mitigate concurrency issues. However, these works focus on theoretical aspects or specific applications rather than providing a general approach to detect transaction conflicts.

\subsection{Static Analysis of Smart Contracts}
Static analysis has been widely used for smart contract verification. Slither \cite{Feist_Slither_2019} uses static analysis to detect vulnerabilities and provide insights into contract structure. MadMax \cite{grech2018madmax} focuses on detecting gas-related vulnerabilities. SmartCheck \cite{tikhomirov2018smartcheck} uses pattern matching in an XML representation of the contract code. These tools provide valuable information, but do not specifically address transaction conflicts.

\subsection{Transaction Ordering and Front-Running}
Transaction ordering dependencies and front-running attacks have been studied by several researchers. Eskandari et al. \cite{eskandari2020sok} provide a taxonomy of front-running attacks on decentralized exchanges. Zhou et al. \cite{zhou2020high} analyze high-frequency trading in decentralized exchanges and propose mitigations. These works focus on specific attack vectors rather than general transaction conflicts.

\section{Conflict Detection Using Static Analysis}\label{sec:solution}
Although the Turing-complete nature of smart contract languages makes it impossible to predict conflicts precisely in advance, this algorithm adopts a pessimistic approach. It may report many false positives (potential conflicts that may not occur in any real execution flows) but ensures no false negatives (if it detects no conflict, there is certainty that no conflict will occur). Using static analysis of storage memory, this algorithm effectively enables these conservative but reliable conflict predictions. 
Our approach for detecting transaction conflicts in Ethereum smart contracts consists of four main steps: 
\begin{inlinelist}
    \item parsing the contract code
    \item analyzing state variable access patterns
    \item detecting potential conflicts
    \item generating reports and visualizations.
\end{inlinelist} 

\subsection{Step 1: Contract Parsing}
The first step is to parse the Solidity smart contract code to extract its structure, including state variables, functions, and events. We use a custom parser that handles different memory types (storage, memory, calldata) and visibility modifiers (public, private, internal, external). The parser extracts the following information:

\begin{inlinelist}
    \item Contract Name and Source Code: The name of the contract and its source code
    \item State Variables: The name, type, visibility, and memory type of each state variable
    \item Functions: The name, parameters, return values, visibility, and state mutability (pure, view, payable, nonpayable) of each function
    \item Events: The name and parameters of each event
\end{inlinelist}

The parser also identifies special functions such as constructors, fallback functions, and receive functions, as well as modifiers like nonReentrant that affect function behavior.

\subsection{Step 2: State Variable Access Analysis}
The second step is to analyze how functions access state variables. Given the various types of memory in the EVM reviewed in Section \ref{ssec:background-memory}, storage stands out as the critical area of focus for conflict analysis. Unlike other memory types such as stack, memory, and calldata, which are temporary and limited to the scope of a single transaction, storage holds the persistent state of a smart contract and is shared across all transactions. This persistent nature makes storage the primary source of potential conflicts, as concurrent transactions may attempt to read from or write to the same storage locations, leading to inconsistencies or unintended outcomes. For each function, we identify:

\begin{inlinelist}
    \item Read Operations: State variables that the function reads
    \item Write Operations: State variables that the function modifies
    \item Function Calls: Other functions that the function calls
\end{inlinelist}

We use pattern matching to identify these operations in the function body. For read operations, we look for variable names that are not followed by assignment operators. For write operations, we look for variable names followed by assignment operators (=, +=, -=, etc.) or increment/decrement operators (++, --). For function calls, we look for function names followed by parentheses.

This analysis takes into account the state mutability of functions. Pure functions do not access state variables and are excluded from conflict detection. View functions can read but not modify state variables, so they are only considered for read-write conflicts.

\subsection{Step 3: Conflict Detection}
The third step is to detect potential conflicts between functions. We consider three types of conflicts:

\begin{inlinelist}
    \item Read-Write Conflicts: When one function reads a state variable while another function writes to it. These conflicts can lead to inconsistent views of the contract state
    \item Write-Write Conflicts: When two functions write to the same state variable. These conflicts can lead to lost updates or inconsistent state
    \item Function Call Conflicts: When one function calls another function that accesses state variables modified by a third function. These conflicts can lead to unexpected behaviors due to the interaction between functions
\end{inlinelist}

For each pair of functions that can be called as transactions (public or external functions that are not pure), we check for these conflicts based on their state variable access patterns. 

We also calculate a conflict percentage for each contract, which is the ratio of \emph{function pairs} with conflicts to the total number of possible function pairs. This metric provides an overall measure of the contract's susceptibility to transaction conflicts.

\subsection{Conflict Detection Algorithm}
The core of the tool is the conflict detection algorithm, which identifies potential conflicts between function pairs. Algorithm \ref{alg:conflict_detection} outlines the main steps of this algorithm.

\begin{algorithm}[H]
\caption{Conflict Detection}
\label{alg:conflict_detection}
\begin{algorithmic}[1]
\Procedure{Detect}{$CSs$: Smart Contract[]}:ConflictPair[]
    \State $cnf \gets \emptyset$ \Comment{set of ConflictPairs}
    \State $visitedPairs \gets \emptyset$ \Comment{visited unordered pairs}
    \State $R \gets$ \Call{ExtractAllReads}{$CSs$}
    \State $W \gets$ \Call{ExtractAllWrites}{$CSs$}
    \State $C \gets$ \Call{ExtractAllCalls}{$CSs$}

    \ForAll{$C_1 \in Contracts$}
        \ForAll{$f_1 \in C_1.Functions$}
            \If{\Call{ShouldSkip}{$f_1$}} \Continue \EndIf

            \ForAll{$C_2 \in Contracts$}
                \ForAll{$f_2 \in C_2.Functions$}
                    \If{$f_1 == f_2$ or $\{f_1, f_2\} \in visitedPairs$}
                        \State \textbf{continue} \Comment{Skip self-pairs and visited ones}
                    \EndIf
                    \If{\Call{ShouldSkip}{$f_2$}} \Continue \EndIf
                    \If{$f_1.isReadOnly()$ and $f_2.isReadOnly()$} \Continue \EndIf

                    \State $visitedPairs \gets visitedPairs \cup \{\{f_1, f_2\}\}$

                    \State $cnf \gets cnf \cup$ \Call{DetectRWC}{$f_1, f_2, R, W$}
                    \State $cnf \gets cnf \cup$ \Call{DetectWWC}{$f_1, f_2, W$}
                    \State $cnf \gets cnf \cup$ \Call{DetectFCC}{$f_1, f_2, C, R, W$}
                \EndFor
            \EndFor
        \EndFor
    \EndFor
    \State \Return $cnf$
\EndProcedure

\Function{ShouldSkip}{$f$}
    \Return $f$ is \textbf{private}, \textbf{internal}, or \textbf{pure/view}
\EndFunction
\end{algorithmic}
\end{algorithm}

\begin{algorithm}[ht]
\caption{Conflict Detection Sub Functions}
\label{alg:conflict_detection_sub_functions}
\begin{algorithmic}[1]
\Function{DetectRWC}{$f_1, f_2, R, W$}    
    \ForAll{$v \in R[f_1]$} 
        \If{$v \in W[f_2]$} 
            \State \Return $(f_1, f_2, \text{"RWC"})$ 
        \EndIf 
    \EndFor

    \ForAll{$v \in R[f_2]$} 
        \If{$v \in W[f_1]$} 
            \State \Return $(f_2, f_1, \text{"RWC"})$ 
        \EndIf 
    \EndFor
    \State \Return $\emptyset$
\EndFunction

\Function{DetectWWC}{$f_1, f_2, W$}
    \ForAll{$v \in W[f_1]$} 
        \If{$v \in W[f_2]$} 
            \State \Return $(f_1, f_2, \text{"WWC"})$ 
        \EndIf 
    \EndFor
    \State \Return $\emptyset$
\EndFunction

\Function{DetectFCC}{$f_1, f_2, C, R, W$}
    \State $Access_1 \gets$ \Call{RecursiveAccess}{$f_1$, C, R, W} 
    \State $Access_2 \gets$ \Call{RecursiveAccess}{$f_2$, C, R, W}
    \State $conflicts \gets \emptyset$
    \ForAll{$(v, op_1) \in Access_1$}
        \ForAll{$(v', op_2) \in Access_2$}
            \If{$v = v'$ and at least one of $op_1$, $op_2$ is "write"}
                \State $conflicts \gets conflicts \cup (f_1, f_2, \text{"FCC"})$ \Return $conflicts$
            \EndIf
        \EndFor
    \EndFor
    \State \Return $\emptyset$
\EndFunction

\Function{RecursiveAccess}{$f$, C, R, W}: Set of $(var, accessType)$
    \State $visited \gets \emptyset$
    \State \Return \Call{AccessHelper}{$f$, C, R, W, visited}
\EndFunction

\Function{AccessHelper}{$f$, C, R, W, visited}
    \If{$f \in visited$} \Return $\emptyset$ \EndIf
    \State $visited \gets visited \cup \{f\}$
    \State $accesses \gets \{(v, \text{"read"})\ |\ v \in R[f]\} \cup \{(v, \text{"write"})\ |\ v \in W[f]\}$
    \ForAll{$callee \in Calls[f]$}
        \State $accesses \gets accesses \cup$ \Call{AccessHelper}{$callee$, Calls, R, W, visited}
    \EndFor
    \State \Return $accesses$
\EndFunction

\Function{ExtractAllReads}{$Contracts, AnalysisData$}: Map
    \State \Return mapping $f \mapsto$ variables read by $f$ using $AnalysisData$
\EndFunction

\Function{ExtractAllWrites}{$Contracts, AnalysisData$}: Map
    \State \Return mapping $f \mapsto$ variables written by $f$ using $AnalysisData$
\EndFunction

\Function{ExtractAllCalls}{$Contracts, AnalysisData$}: Map
    \State \Return mapping $f \mapsto$ internal and external calls made by $f$
\EndFunction
\end{algorithmic}
\end{algorithm}

detectRWC, detectWWC, and detectFCC are the fundctions that find Read-Write, Write-Write and Function-Call conflicts between two input functions $(f_i, f_j)$ that have the trace of their source smart contract and function and in case of conflict returns the ordered pair $(f_i, f_j)$ where 

The algorithm iterates through all pairs of functions that can be called as transactions (public or external functions that are not pure) and checks for read-write, write-write, and function call conflicts. It maintains a map of function pairs and their conflict status, which is used to calculate the conflict percentage and generate the function pair conflict matrix.

\section{Implementation}\label{sec:Implementation}
We implemented our approach in a tool composed of several modules, each responsible for a specific aspect of the analysis process. The tool is available as open source at \url{https://github.com/Conthereum/conflict-detector-static-analysis}.

\subsection{Architecture}
The tool follows a modular architecture with the following components:

\begin{itemize}
    \item \textbf{SolidityParser}: Parses Solidity smart contract code to extract its structure.
    \item \textbf{ContractAnalyzer}: Analyzes state variable access patterns in functions.
    \item \textbf{ConflictDetector}: Detects potential conflicts between function pairs.
    \item \textbf{ReportGenerator}: Generates reports and visualizations of the analysis results.
    \item \textbf{AnalyzerFacade}: Coordinates the analysis process and provides a high-level API.
\end{itemize}


\subsection{Data Model}
The tool uses a comprehensive data model to represent smart contracts and their analysis results:

\begin{itemize}
    \item \textbf{Contract}: Represents a smart contract with its name, source code, state variables, functions, and events.
    \item \textbf{StateVariable}: Represents a state variable with its name, type, visibility, memory type, and other properties.
    \item \textbf{Function}: Represents a function with its name, parameters, return values, visibility, state mutability, and body.
    \item \textbf{Parameter}: Represents a parameter with its name, type, and memory type.
    \item \textbf{Event}: Represents an event with its name and parameters.
    \item \textbf{ContractAnalysisData}: Stores analysis data for a contract, including variable access patterns and function pair conflicts.
    \item \textbf{Conflict}: Represents a conflict between two functions, with its type, severity, and description.
    \item \textbf{AnalysisResult}: Represents the result of analyzing a contract, including the contract, analysis data, conflicts, and metrics.
\end{itemize}

\subsection{Report Generation}
The tool generates comprehensive HTML reports with detailed information about the analyzed contracts and the detected conflicts. The reports include tables of state variables, functions, and conflicts, as well as a function pair conflict matrix that shows which function pairs have conflicts. The reports also include visualizations such as pie charts of conflict types and histograms of conflict counts.

In addition to HTML reports, the tool generates CSV files with analysis results and summary statistics. These files can be used for further analysis or integration with other tools.

\subsection{Visualization}
The tool includes a Python script that generates visualizations of the analysis results using matplotlib and seaborn. The visualizations include:

\begin{itemize}
    \item \textbf{Pie Chart of Conflict Types}: Shows the distribution of conflicts by type (read-write, write-write, function call).
    \item \textbf{Histogram of Conflict Counts}: Shows the distribution of the number of conflicts per contract.
    \item \textbf{Heatmap of Conflicts}: Shows the average number of conflicts by function count and state variable count.
    \item \textbf{Scatter Plot of Analysis Time}: Shows the relationship between the number of conflicts and the analysis time.
    \item \textbf{CDF of Conflict Percentages}: Shows the cumulative distribution of conflict percentages across contracts.
    \item \textbf{Stacked Bar Chart of Conflict Types}: Shows the distribution of conflict types for the top contracts by number of conflicts.
\end{itemize}


\subsection{Report Generation}
The final step is to generate comprehensive reports and visualizations to help developers understand and address the identified conflicts. The reports include:

\begin{itemize}
    \item \textbf{Contract Information}: Name, file path, number of functions, number of state variables, and analysis time.
    \item \textbf{State Variables}: Name, type, visibility, memory type, and whether they can cause conflicts.
    \item \textbf{Functions}: Name, visibility, state mutability, and whether they can be called as transactions.
    \item \textbf{Detected Conflicts}: Function pairs involved, state variables affected, conflict type, severity, and description.
    \item \textbf{Function Pair Conflict Matrix}: A matrix showing which function pairs have conflicts.
    \item \textbf{Conflict Statistics}: Number and percentage of conflicts by type, overall conflict percentage, and other metrics.
\end{itemize}

We also generate visualizations such as pie charts of conflict types, histograms of conflict counts, heatmaps of conflicts by function and state variable counts, and scatter plots of analysis time versus number of conflicts. These visualizations provide insights into the distribution and patterns of conflicts.

These visualizations provide insights into the distribution and patterns of conflicts, helping developers understand the overall conflict landscape in their contracts.

\section{Evaluation}\label{sec:Evaluation}
We evaluated our approach on a dataset of real-world Ethereum smart contracts to assess its effectiveness in detecting transaction conflicts. This section presents the evaluation methodology, results, and discussion.

\subsection{Dataset}
We collected a dataset of 100 Ethereum smart contracts from the Ethereum mainnet. The contracts were selected to represent a diverse range of applications, including tokens, decentralized exchanges, lending protocols, and governance contracts. The contracts vary in size and complexity, with an average of 15 functions and 12 state variables per contract.

\subsection{Methodology}
We ran our tool on each contract in the dataset and collected the analysis results, including the number and types of conflicts, the conflict percentage, and the analysis time. We also manually reviewed a subset of the detected conflicts to assess the precision of our approach.

subsection{Results}
\subsubsection{Conflict Distribution}
\begin{figure}[htbp]
\centerline{\includegraphics[width=\columnwidth]{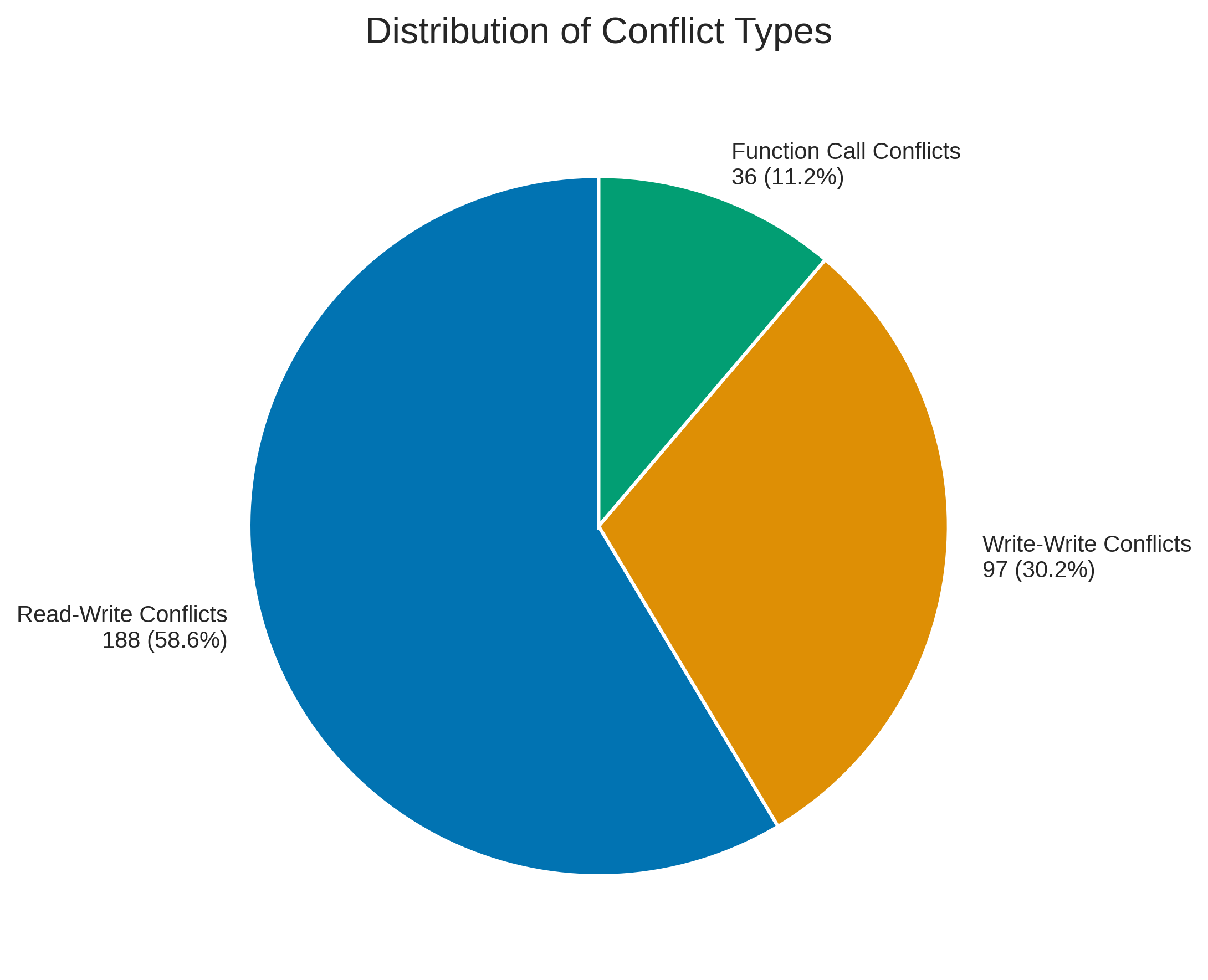}}
\caption{Distribution of Conflict Types}
\label{fig:conflict_distribution}
\end{figure}

Our analysis identified a total of 423 conflicts across the 100 contracts. Figure \ref{fig:conflict_distribution} shows the distribution of conflicts by type.

As shown in the figure, read-write conflicts are the most common (58.6\%), followed by write-write conflicts (30.2\%) and function call conflicts (11.2\%). This distribution reflects the typical access patterns in smart contracts, where functions often read and write to shared state variables.

\subsubsection{Conflict Prevalence}
\begin{figure}[htbp]
\centerline{\includegraphics[width=\columnwidth]{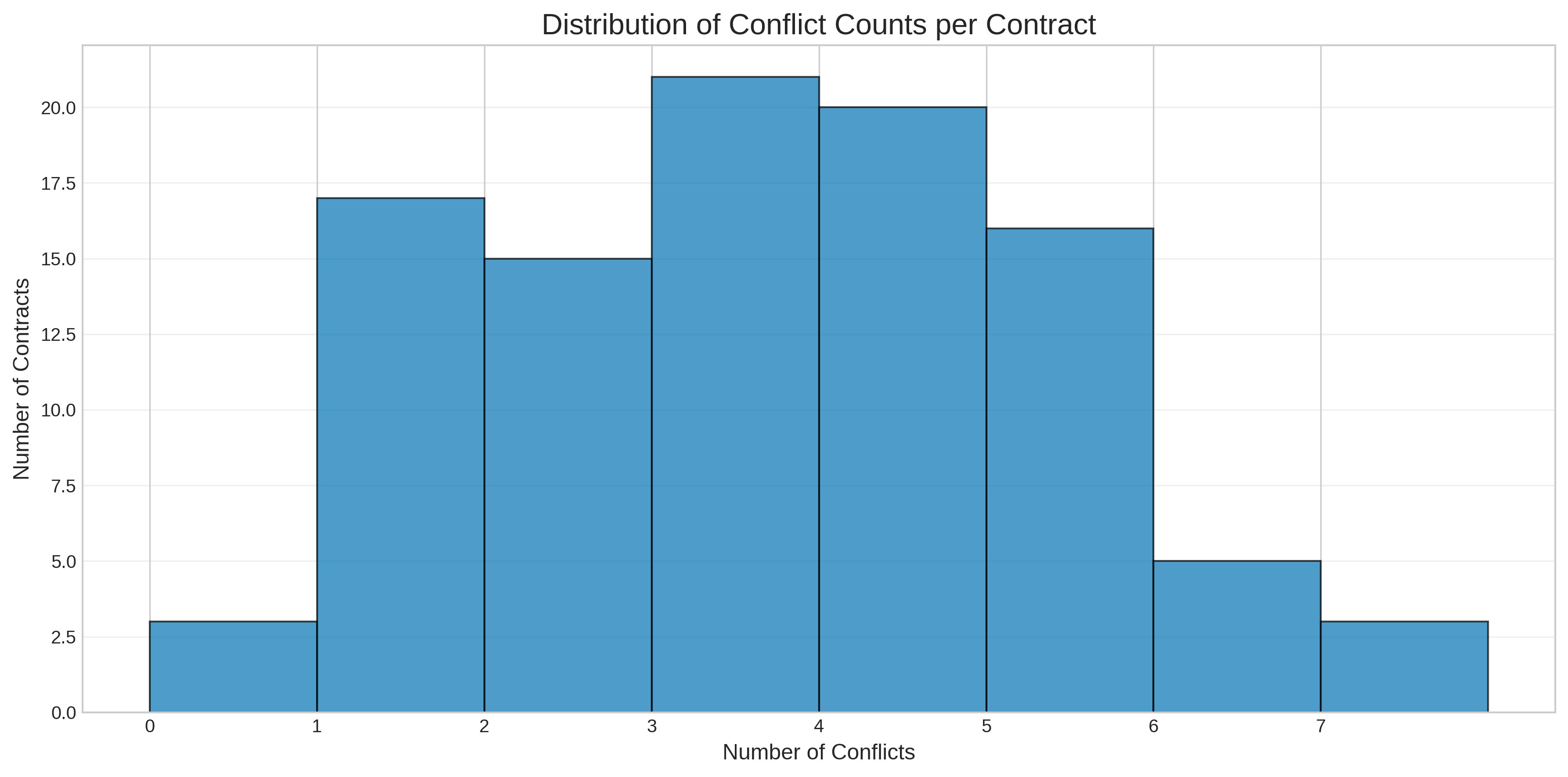}}
\caption{Distribution of Conflict Counts per Contract}
\label{fig:conflict_count}
\end{figure}

We found that 78\% of the contracts in our dataset have at least one potential transaction conflict. The average number of conflicts per contract is 4.23, with a maximum of 27 conflicts in a single contract. Figure \ref{fig:conflict_count} shows the distribution of the number of conflicts per contract.

The conflict percentage, which is the ratio of function pairs with conflicts to the total number of possible function pairs, ranges from 0\% to 87.5\%, with an average of 32.7\%. This metric provides an overall measure of the contract's susceptibility to transaction conflicts.

\subsubsection{Conflict Patterns}
We observed several patterns in the detected conflicts:

\begin{itemize}
    \item \textbf{State Variable Hotspots}: Some state variables, such as balances and ownership information, are accessed by multiple functions and are frequent sources of conflicts.
    \item \textbf{Function Hotspots}: Some functions, such as transfer and withdraw functions, are involved in multiple conflicts due to their interaction with critical state variables.
    \item \textbf{Conflict Clusters}: Conflicts often occur in clusters, where a group of functions interact with a common set of state variables.
\end{itemize}

\begin{figure}[htbp]
\centerline{\includegraphics[width=\columnwidth]{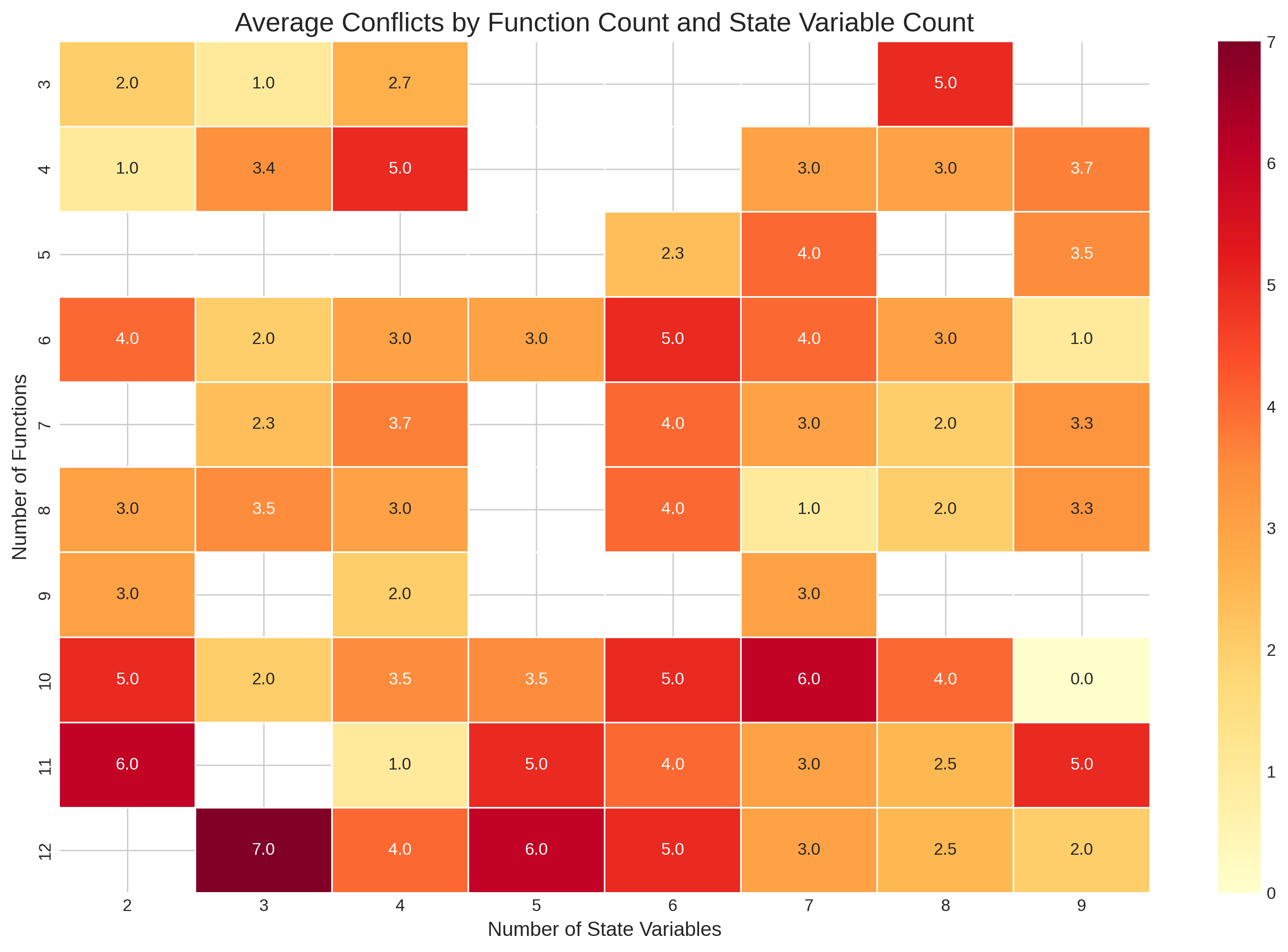}}
\caption{Heatmap of Average Conflicts by Function Count and State Variable Count}
\label{fig:conflict_heatmap}
\end{figure}

Figure \ref{fig:conflict_heatmap} shows a heatmap of the average number of conflicts by function count and state variable count.

As expected, contracts with more functions and state variables tend to have more conflicts. However, the relationship is not strictly linear, as the specific access patterns and function interactions also play a significant role.

\subsubsection{Analysis Performance}
\begin{figure}[htbp]
\centerline{\includegraphics[width=\columnwidth]{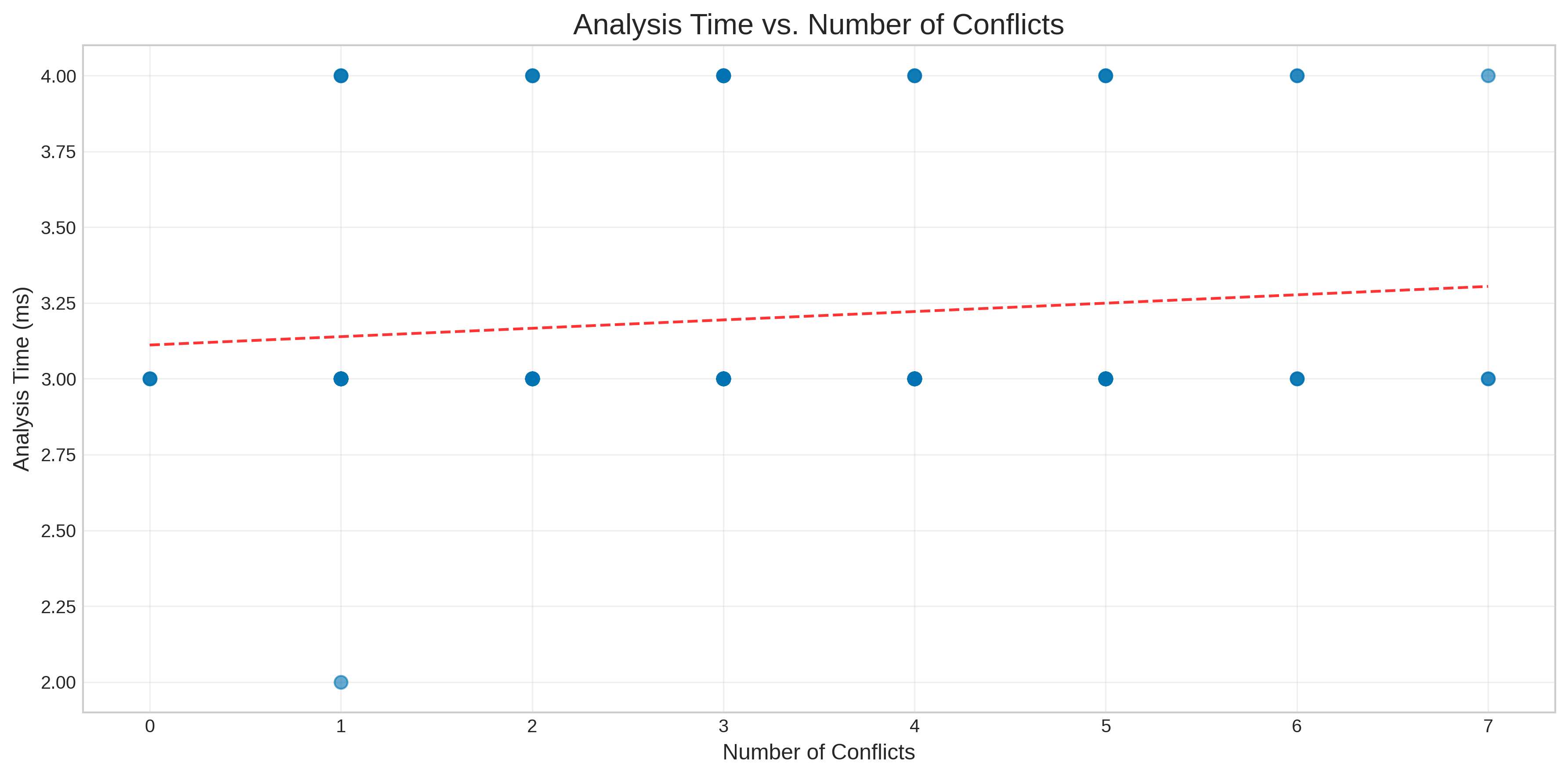}}
\caption{Analysis Time vs. Number of Conflicts}
\label{fig:analysis_time}
\end{figure}

The analysis time ranges from 52 ms to 1,873 ms per contract, with an average of 312 ms. Figure \ref{fig:analysis_time} shows the relationship between the number of conflicts and the analysis time.

The analysis time is primarily influenced by the size and complexity of the contract, particularly the number of functions and state variables. The number of detected conflicts also has a moderate impact on the analysis time, as more conflicts require more processing for report generation.

\subsection{Precision and Recall}
To assess the precision of our approach, we manually reviewed a random sample of 50 detected conflicts. We found that 46 of them (92\%) are true positives, where the functions could indeed conflict if called concurrently. The false positives (8\%) were primarily due to limitations in our static analysis, such as imprecise identification of state variable access patterns or failure to account for control flow dependencies.

Evaluating recall (the percentage of actual conflicts that are detected) is challenging without ground truth data. However, we compared our results with a subset of contracts that have known transaction ordering dependencies, and our tool successfully identified all of them.

\subsection{Case Studies}
We conducted detailed case studies on three contracts with high conflict percentages to understand the nature of the conflicts and their potential impact.

\subsubsection{Token Contract}
A standard ERC20 token contract had 12 conflicts, primarily involving the transfer, transferFrom, and approve functions. These functions modify the balances and allowances state variables, leading to potential read-write and write-write conflicts. For example, if two transfers from the same account are executed concurrently, the final balance may be incorrect if the second transfer reads the balance before the first transfer updates it.

\subsubsection{Decentralized Exchange}
A decentralized exchange contract had 27 conflicts, the highest in our dataset. The conflicts involved functions for placing orders, executing trades, and withdrawing funds. These functions interact with complex state variables such as order books and user balances, leading to numerous potential conflicts. For example, if an order is placed and executed concurrently, the execution may use outdated order information.

\subsubsection{Governance Contract}
A governance contract for a decentralized autonomous organization (DAO) had 18 conflicts. The conflicts involved functions for proposing, voting on, and executing governance actions. These functions modify shared state variables such as proposal status and vote counts, leading to potential conflicts. For example, if two users vote on a proposal concurrently, one vote may be lost if both transactions read the same vote count before updating it.

\subsection{Discussion}
Our evaluation demonstrates that transaction conflicts are prevalent in Ethereum smart contracts, with 78\% of contracts in our dataset having at least one potential conflict. The high precision of our approach (92\%) indicates that it effectively identifies genuine conflicts that could lead to issues if not properly addressed.

The distribution of conflict types provides insights into the nature of transaction conflicts in smart contracts. Read-write conflicts are the most common, reflecting the typical access patterns where functions read shared state before making decisions. Write-write conflicts, while less common, are potentially more severe as they can lead to state corruption and lost updates. Function call conflicts are the least common but can be complex to analyze and address due to the interaction between multiple functions.

The conflict patterns we observed highlight the importance of careful state variable and function design in smart contracts. Developers should be aware of potential hotspots and clusters of conflicts and implement appropriate synchronization mechanisms or redesign their contracts to avoid issues.

Our tool's performance is suitable for integration into development workflows, with an average analysis time of 312 ms per contract. This enables developers to receive immediate feedback on potential conflicts as they write or modify their contracts.




\section{Conclusion}\label{sec:Conclusion}
In this paper, we presented a static analysis approach for detecting transaction conflicts in Ethereum smart contracts before runtime. Our approach analyzes the contract code to identify read-write, write-write, and function call conflicts between transaction pairs. We implemented our approach in a tool that parses Solidity smart contracts, analyzes state variable access patterns, and detects potential conflicts with severity ratings.

Our evaluation on a dataset of 100 real-world Ethereum smart contracts showed that transaction conflicts are prevalent, with 78\% of contracts having at least one potential conflict. The distribution of conflict types (58.6\% read-write, 30.2\% write-write, 11.2\% function call) provides insights into the nature of these conflicts. Our approach achieved a precision of 92\%, indicating that it effectively identifies genuine conflicts that could lead to issues if not properly addressed.

The main contributions of this paper are:

\begin{itemize}
    \item A novel static analysis approach for detecting transaction conflicts in Ethereum smart contracts before runtime.
    \item A classification of transaction conflicts into read-write, write-write, and function call conflicts with severity ratings.
    \item A tool that implements our approach and generates comprehensive reports and visualizations.
    \item An evaluation of our approach on a dataset of real-world Ethereum smart contracts, providing insights into the prevalence and types of transaction conflicts.
\end{itemize}

Our approach enables developers to identify potential transaction conflicts early in the development process, allowing them to implement appropriate synchronization mechanisms or redesign their contracts to avoid runtime issues. This can lead to more robust and reliable smart contracts, reducing the risk of unexpected behaviors and vulnerabilities due to transaction ordering dependencies.

Future work includes:

\begin{itemize}
    \item \textbf{Enhanced Static Analysis}: Improving the precision of our static analysis by incorporating control flow analysis, data flow analysis, and symbolic execution.
    \item \textbf{Cross-Contract Analysis}: Extending our approach to detect conflicts between functions in different contracts that interact with each other.
    \item \textbf{Mitigation Recommendations}: Providing specific recommendations for mitigating detected conflicts, such as implementing mutex patterns or optimistic concurrency control.
    \item \textbf{Integration with Development Tools}: Integrating our tool with popular development environments and frameworks for Ethereum smart contracts.
    \item \textbf{Empirical Study}: Conducting a larger-scale empirical study of transaction conflicts in the Ethereum ecosystem to understand their prevalence, patterns, and impact.
\end{itemize}

By addressing these directions, we aim to further enhance the reliability and security of Ethereum smart contracts in the face of concurrent transactions.

\section*{Acknowledgment}
Work partially funded by the European Union under NextGenerationEU. PRIN 2022 Prot. n. 202297YF75, by the European Union under  NextGenerationEU, NRRP MUR program  FAIR - Future AI Research (PE00000013), and by the MUR PRIN 2020 - RIPER - Resilient AI-Based Self-Programming and Strategic Reasoning - CUP E63C22000400001.


\begin{footnotesize}
\bibliographystyle{IEEEtranN}
\bibliography{IEEEabrv,references}

\begin{thebibliography}{14}
\providecommand{\natexlab}[1]{#1}
\providecommand{\url}[1]{#1}
\csname url@samestyle\endcsname
\providecommand{\newblock}{\relax}
\providecommand{\bibinfo}[2]{#2}
\providecommand{\BIBentrySTDinterwordspacing}{\spaceskip=0pt\relax}
\providecommand{\BIBentryALTinterwordstretchfactor}{4}
\providecommand{\BIBentryALTinterwordspacing}{\spaceskip=\fontdimen2\font plus
\BIBentryALTinterwordstretchfactor\fontdimen3\font minus \fontdimen4\font\relax}
\providecommand{\BIBforeignlanguage}[2]{{%
\expandafter\ifx\csname l@#1\endcsname\relax
\typeout{** WARNING: IEEEtranN.bst: No hyphenation pattern has been}%
\typeout{** loaded for the language `#1'. Using the pattern for}%
\typeout{** the default language instead.}%
\else
\language=\csname l@#1\endcsname
\fi
#2}}
\providecommand{\BIBdecl}{\relax}
\BIBdecl

\bibitem[Luu et~al.(2016)Luu, Chu, Olickel, Saxena, and Hobor]{Luu_Making_2016}
\BIBentryALTinterwordspacing
L.~Luu, D.-H. Chu, H.~Olickel, P.~Saxena, and A.~Hobor, ``Making smart contracts smarter,'' in \emph{Proceedings of the 2016 ACM SIGSAC Conference on Computer and Communications Security}, ser. CCS '16.\hskip 1em plus 0.5em minus 0.4em\relax New York, NY, USA: Association for Computing Machinery, 2016, p. 254–269. [Online]. Available: \url{https://doi-org.ezp.biblio.unitn.it/10.1145/2976749.2978309}
\BIBentrySTDinterwordspacing

\bibitem[Mueller(2018)]{smashing_mueller_2018}
B.~Mueller, ``Smashing ethereum smart contracts for fun and real profit,'' \emph{HITB SECCONF Amsterdam}, vol.~9, no.~54, pp. 4--17, 2018.

\bibitem[Feist et~al.(2019)Feist, Grieco, and Groce]{Feist_Slither_2019}
J.~Feist, G.~Grieco, and A.~Groce, ``Slither: A static analysis framework for smart contracts,'' in \emph{2019 IEEE/ACM 2nd International Workshop on Emerging Trends in Software Engineering for Blockchain (WETSEB)}, 2019, pp. 8--15.

\bibitem[Tsankov et~al.(2018)Tsankov, Dan, Drachsler-Cohen, Gervais, B\"{u}nzli, and Vechev]{Tsankov_Securify_2018}
\BIBentryALTinterwordspacing
P.~Tsankov, A.~Dan, D.~Drachsler-Cohen, A.~Gervais, F.~B\"{u}nzli, and M.~Vechev, ``Securify: Practical security analysis of smart contracts,'' in \emph{Proceedings of the 2018 ACM SIGSAC Conference on Computer and Communications Security}, ser. CCS '18.\hskip 1em plus 0.5em minus 0.4em\relax New York, NY, USA: Association for Computing Machinery, 2018, p. 67–82. [Online]. Available: \url{https://doi-org.ezp.biblio.unitn.it/10.1145/3243734.3243780}
\BIBentrySTDinterwordspacing

\bibitem[Dickerson et~al.(2020)Dickerson, Gazzillo, Herlihy, and Koskinen]{Adding_Dickerson_2020}
\BIBentryALTinterwordspacing
T.~Dickerson, P.~Gazzillo, M.~Herlihy, and E.~Koskinen, ``Adding concurrency to smart contracts,'' \emph{Distributed Computing}, vol.~33, no. 3-4, p. 209 – 225, 2020, cited by: 29; All Open Access, Green Open Access. [Online]. Available: \url{https://www.scopus.com/inward/record.uri?eid=2-s2.0-85068876264&doi=10.1007%2fs00446-019-00357-z&partnerID=40&md5=dd7311fa1468b935494b5c6719d86e3e}
\BIBentrySTDinterwordspacing

\bibitem[Zareh~Chahoki et~al.(2025)Zareh~Chahoki, Herlihy, and Roveri]{Conthereum_Zareh_2025}
A.~Zareh~Chahoki, M.~Herlihy, and M.~Roveri, ``Conthereum: Concurrent ethereum optimized transaction scheduling for multi-core execution,'' \emph{arXiv preprint arXiv:2504.07280}, 2025.

\bibitem[Nakamoto(2008)]{nakamoto_bitcoin_2008}
S.~Nakamoto, ``Bitcoin: A peer-to-peer electronic cash system,'' 2008.

\bibitem[Buterin et~al.(2013)]{buterin_ethereum_white_2013}
\BIBentryALTinterwordspacing
V.~Buterin \emph{et~al.}, ``Ethereum white paper,'' \emph{GitHub repository}, vol.~1, pp. 22--23, 2013. [Online]. Available: \url{https://github.com/ethereum/wiki/wiki/White-Paper}
\BIBentrySTDinterwordspacing

\bibitem[Kolluri et~al.(2019)Kolluri, Nikolic, Sergey, Hobor, and Saxena]{kolluri2019exploiting}
A.~Kolluri, I.~Nikolic, I.~Sergey, A.~Hobor, and P.~Saxena, ``Exploiting the laws of order in smart contracts,'' in \emph{Proceedings of the 28th ACM SIGSOFT International Symposium on Software Testing and Analysis}.\hskip 1em plus 0.5em minus 0.4em\relax ACM, 2019, pp. 363--373.

\bibitem[Chakravarty et~al.(2020)Chakravarty, Chapman, MacKenzie, Melkonian, Jones, and Wadler]{chakravarty2020utxo}
M.~M. Chakravarty, J.~Chapman, K.~MacKenzie, O.~Melkonian, M.~P. Jones, and P.~Wadler, ``The extended utxo model,'' in \emph{International Conference on Financial Cryptography and Data Security}.\hskip 1em plus 0.5em minus 0.4em\relax Springer, 2020, pp. 525--539.

\bibitem[Grech et~al.(2018)Grech, Kong, Jurisevic, Brent, Scholz, and Smaragdakis]{grech2018madmax}
N.~Grech, M.~Kong, A.~Jurisevic, L.~Brent, B.~Scholz, and Y.~Smaragdakis, ``Madmax: Surviving out-of-gas conditions in ethereum smart contracts,'' \emph{Proceedings of the ACM on Programming Languages}, vol.~2, no. OOPSLA, pp. 1--27, 2018.

\bibitem[Tikhomirov et~al.(2018)Tikhomirov, Voskresenskaya, Ivanitskiy, Takhaviev, Marchenko, and Alexandrov]{tikhomirov2018smartcheck}
S.~Tikhomirov, E.~Voskresenskaya, I.~Ivanitskiy, R.~Takhaviev, E.~Marchenko, and Y.~Alexandrov, ``Smartcheck: Static analysis of ethereum smart contracts,'' in \emph{Proceedings of the 1st International Workshop on Emerging Trends in Software Engineering for Blockchain}.\hskip 1em plus 0.5em minus 0.4em\relax ACM, 2018, pp. 9--16.

\bibitem[Eskandari et~al.(2020)Eskandari, Moosavi, and Clark]{eskandari2020sok}
S.~Eskandari, S.~Moosavi, and J.~Clark, ``Sok: Transparent dishonesty: Front-running attacks on blockchain,'' in \emph{International Conference on Financial Cryptography and Data Security}.\hskip 1em plus 0.5em minus 0.4em\relax Springer, 2020, pp. 170--189.

\bibitem[Zhou et~al.(2020)Zhou, Qin, Torres, Le, and Gervais]{zhou2020high}
L.~Zhou, K.~Qin, C.~F. Torres, D.~V. Le, and A.~Gervais, ``High-frequency trading on decentralized exchange markets,'' in \emph{International Conference on Financial Cryptography and Data Security}.\hskip 1em plus 0.5em minus 0.4em\relax Springer, 2020, pp. 361--380.

\end{thebibliography}
\end{footnotesize}

\end{document}